\documentclass[prb,aps,preprint,superscriptaddress]{revtex4}
\usepackage{graphicx}
\usepackage{amssymb}
\usepackage{bm}

\begin{document}

\author{Toshifumi Itakura}
 \affiliation{ASPRO TEC., Takabaridai 3-506, Meito-ku, Nagoya, 465-0054, Japan}
 \email{itakurat@s6.dion.ne.jp}

\title{Decoherence of coupled qubit system}

\begin{abstract}
In this study, we examine the decoherence 
of independently coupled qubit systems by using the influence functional.
We especially concentrated on the effect of qubit flip process.
We examine the zero-dimensional qubit and one-dimensional qubit systems coupled with the multi-band one-dimensional system.
The qubit flip process leads to self-excitation and oscillation.
We obtained the time evolution of von Neumann entropy.
We also estimated
  numerically time evolution of density matrix and the entropy.
One of the results indicates that the slower random telegraph noise leads to the more mixed states.
Thus the background charge fluctuation is important for charge qubit.

\end{abstract}


\maketitle

\section{Introduction}
Among various proposals for quantum computations, quantum bits (qubits) in solid state materials, such as superconducting Josephson junctions
\cite{Nakamura},
           and quantum dots
\cite{Hayashi,Tanamoto,Loss},
           have the advantage of  scalability.
Such coherent two level systems constitute qubits and
 a quantum computation can be carried out as a unitary operation applied to many qubit systems.
It is essential that this quantum coherence is maintained during computation.
However, dephasing is a hard to avoid, due to the interaction between qubit system and the environment.
The decay of off-diagonal elements of the qubit density matrix shows dephasing.
Various environments can cause dephasing.
The effect of phonons is important in solid systems, 
\cite{Fujisawa_SC}.
The effect of electromagnetic fluctuation
     has been extensively studied for Josephson junction charge qubits
\cite{Schon}.
The background charge fluctuation is important for charge qubit system.
\cite{Itakura_Tokura}

The exact expression of Green's functions are obtained for a free fermion and boson system and multi-band system.
\cite{Zagoskin,Itakura_Kawakami}
By using these expressions,
we consider  that qubit systems is coupled to multi-band one-dimensional bath. 
Integrating over the multi-band one-dimensional system's variables,
we obtain the qubit density matrix.
\cite{Feynman,Weiss,Itakura}
Using this influence functional method,
we examine the dephasing of the qubit density matrix.
In the present study, we concentrated on the effect of qubit flip process.
The qubit flip process lead to self-excitation and oscillation.
\cite{Tokura_s}
We obtained the time evolution of von Neumann entropy.
We also present the numerical calculation of the density matrix and entropy.
One of the results indicates that the slower random telegraph noise 
leads to the more mixed states.
Thus the background charge fluctuation is important for charge qubit.

\section{Hamiltonian}
We examine the Hamiltonian of multi-band 
fermion or system
\cite{Itakura_Kawakami}
and qubit systems. 
The type of bath and qubit interaction is XY,
\begin{eqnarray}
H_{bath} &=&  
 [ 
        \sum_{i=1}^N [\sum_{\sigma=\uparrow,\downarrow} \sum_{m=1}^2
       -t ( c_{i \sigma}^{m \dagger} c_{i+1 \sigma}^m 
       + c_{i+1 \sigma}^{m \dagger} c_{i \sigma}^m)  \nonumber \\
    &+&   J ( c^{m \dagger}_{i \sigma} c^{m'}_{i \sigma'}
                       c^{m' \dagger}_{i+1 \sigma'} c^m_{i+1 \sigma}
                       - n_i n_{i+1} )]  \nonumber \\
                       &-&\Delta_B ( n_i^{(1)} -n_i^{(2)} )] \\
H_{qubit} &=& \sum_i \hbar \omega_{oi} I^{z,i} \\
H_{int} &=& \sum_{i,\alpha,m} \gamma_N \hbar^2 
A_{\perp} ( I^{+}_i c_{i,\alpha}^m 
 + I^{-}_i c^{m \dagger}_{i,\alpha} ) 
\end{eqnarray}
$c_{i \sigma}^{m \dagger}$,$c_{i \sigma}^m$ 
are bath creation and destruction 
operator at site $i$ of one-dimensional chain.
$\sigma$ is spin index and m is multi-band index.
For canonical commutation relation is given by
\begin{eqnarray}
c^{\dagger,m}_{i,\sigma}  c_{j,\sigma'}^{m'}
 \pm c_{j,\sigma'}^{m'} c^{\dagger,m}_{j,\sigma} 
 &=& \delta_{i,j} \delta_{\sigma,\sigma'} \delta_{m,m'}, \nonumber \\
c^{\dagger,m}_{i,\sigma} c^{\dagger,m'}_{j,\sigma'}
 \pm c^{\dagger,m'}_{j,\sigma'}
 c^{\dagger,m}_{i,\sigma'} &=& 0, \nonumber \\
c^m_{i,\sigma} c^{m'}_{j,\sigma'} \pm c^{m'}_{j,\sigma'} c^m_{i,\sigma} &=& 0.
\end{eqnarray}
The upper sign is fermion and down sign is boson.
$I^{j}_i$,($j=x,y,z$) is qubit operator at site $i$,$\gamma_N$ is geomagnetic ratio of qubit.
Above model, the one-dimensional bath and zero-dimensional qubit system 
interacts with contact interaction.
We also examine the one-dimensional qubit system where one-dimensional 
bath interact, the effect of indirect interaction appears.

\section{single qubit system}

First, we consider a single qubit and one-dimensional chain system.
Thus, we examine the effect of direct interaction 
by influence functional method.
\cite{Feynman,Weiss,Itakura}
The interaction Hamiltonian is as below,
\begin{equation}
  H_{int} = \gamma_N \hbar^2 [
   A_{\perp} ( I^+_0 c_0+ I^-_0 c_0^{\dagger} )].
\end{equation}
We integrate about the one dimension bath system.
This lead density matrix of qubit system,
\begin{eqnarray}
 \rho ( I_{z+}^f, I_{z-}^f )
   &=& \int^{I_{z+}(t) =I_{z+}^f, I_{z-} (t) =
    I_{z-}^f}_{ I_{z+}(0)=I_{z+}^f, I_{z+} (0) = I_{z+}^f} 
 [ d I_{z+} ] [ d I_{z-} ] \nonumber \\
  && {\exp} ( \frac{i}{\hbar} (I_{qb} [ I_{z-} ] - I_{qb} [ I_{z+} ] ))
  F[I_{z+},I_{z-}], \nonumber \\
\end{eqnarray}
the influence functional is
\begin{eqnarray}
   \label{eqn:inf}
 F[ I_{z+}, I_{z-} ] &=&
  \int \Pi_i [ d c_{+i}^{\dagger}  ] [ d  c_{-i}^{\dagger} ] 
[ d c_{+i}  ] [ d  c_{-i} ] \nonumber \\
 && \delta (  c_{+i} (t) - c_{-i}(t) ) 
 \rho (  c_i  (0),  c_j (0) ) \nonumber \\ 
 && {\rm exp} \{ 
 \frac{i}{\hbar} (  I[c_+ ]- I [c_- ] ) 
\},  
\end{eqnarray}
where, $I_{qb} [I] = \int_0^t \hbar \Delta I_z$.
For following discussion, we choose the action of system,
$ I[ c ] = I_0 [ c ] + I_{int} [c, S_{zi}^0 ]$, 
thus the unperturbed is given by,
\begin{eqnarray}
   I_0 [ c ] &=& \frac{ (A_{\perp} \gamma_N \hbar)^2}{4} 
   \int_0^t  \int_0^{t_1} dt_1 dt_2   \nonumber \\
   && ( c_i^{\dagger} (t_1)
    {\bf \Delta}_{00p}^{+- -1} ( t_1 , i , t_2 , i) 
    c_i ( t_2 ) \nonumber \\
    &+& c_i (t_1)
    {\bf \Delta}_{00p}^{-+ -1} ( t_1 , i , t_2 , i) 
    c_i^{\dagger} ( t_2 )  ), 
\end{eqnarray}
where ${\bf \Delta}_{00p} (t_1,i,t_2, j) $ is the propagator of environmental system which is defined on closed time path and has four components
\begin{eqnarray}
  \label{eqn:Green}
  && {\bf \Delta}_{00p}  (i, t_1, j, t_2) \nonumber \\ 
  &=& \left(
   \begin{array}{cc}
  {\bf \Delta}_{00}^{++} (i, t_1,j, t_2) & {\bf \Delta}_{00}^{+-}  
  (i, t_1, j, t_2) \\
  {\bf \Delta}_{00}^{-+}  (i, t_1,j, t_2) &{\bf  \Delta}_{00}^{--} 
  (i, t_1, j, t_2) 
  \end{array}
  \right). \nonumber \\
\end{eqnarray}
Introducing the incoming interaction picture for the environment system, we can easily verify that Eq. (\ref{eqn:inf}) turns out,
\cite{Chou}
\begin{eqnarray}
  F[ I_{z+}, I_{z-}] &=& 
   {\rm exp} [ - i \frac{(\gamma_N \hbar A_{\perp} )^2}{4} 
    \int_{0}^{t} \int_{0}^{t_1} dt_1 dt_2 \nonumber \\ 
    &&   
   ( I_{++} (t_1) \Delta_{00}^{++} (i,t_1,i,t_2) I_{++} (t_2) \nonumber \\
    &+& I_{+-} (t_1) \Delta_{00}^{--} (i,t_1,i,t_2) I_{+-} (t_2) \nonumber \\
    &-& I_{++} (t_1) \Delta_{00}^{+-} (i,t_1,i,t_2) I_{+-} (t_2) \nonumber \\
    &-& I_{+-} (t_1) \Delta_{00}^{-+} (i,t_1,i,t_2) I_{++} (t_2) )\nonumber \\
  && 
   ( I_{-+} (t_1) \Delta_{00}^{++} (i,t_1,i,t_2) I_{-+} (t_2) \nonumber \\
  &+& I_{--} (t_1) \Delta_{00}^{--} (i,t_1,i,t_2) I_{--} (t_2) \nonumber \\
   &-&  I_{-+} (t_1) \Delta_{00}^{+-} (i,t_1,i,t_2) I_{--} (t_2) \nonumber \\
   &-& I_{--} (t_1) \Delta_{00}^{-+} (i,t_1,i,t_2) I_{-+} (t_2)) ]. \nonumber \\ \end{eqnarray}
For the convenience we change of the coordinate,
\begin{equation}
  \eta_z = (I_{z+} + I_{z-})/2, 
  \xi_z = (I_{z+} - I_{z-})/2.
\end{equation}
The $\eta$ and $\xi$ are called sojourn and blip.
In terms of this variable, density matrix is described by as follows
\begin{eqnarray}
 \rho (\eta_z =1) &=& |\uparrow><\uparrow|,
 \rho ( \eta_z =-1) = |\downarrow><\downarrow|, \nonumber \\
 \rho (\xi_z  = 1 ) &=&  |\uparrow><\downarrow|,
 \rho ( \xi_z = - 1)  = |\downarrow><\uparrow |. \nonumber 
\end{eqnarray}
Other variables are defined by
\begin{equation}
  \eta_+ = (I_{++} + I_{+-})/2, 
  \xi_+ = (I_{++} - I_{+-})/2,
\end{equation}
\begin{equation}
  \eta_- = (I_{-+} + I_{--})/2, 
  \xi_- = (I_{-+} - I_{--})/2.
\end{equation}
In terms of above variables the elements of density matrix are expressed as
\begin{eqnarray}
 \rho (\eta_+ =1) &=& |\uparrow><\downarrow|,
 \rho ( \eta_+ =-1) = |\downarrow><\uparrow|, \nonumber \\
 \rho (\xi_+  = 1 ) &=&  |\uparrow><\uparrow|,
 \rho ( \xi_+ = - 1)  = |\downarrow><\downarrow |. \nonumber \\
 \rho (\eta_- =1 ) &=& |\downarrow><\uparrow|,
 \rho ( \eta_- =-1) = |\uparrow><\downarrow|, \nonumber \\
 \rho (\xi_-  = 1 ) &=&  |\downarrow><\downarrow|,
 \rho ( \xi_- = - 1)  = |\uparrow><\uparrow |. \nonumber 
\end{eqnarray}
Therefore, the below equations are hold for these new variables,
\begin{equation}
\eta_z = \xi_+ = -\xi_- , \xi_z = \eta_+ =-\eta_-.
\end{equation}

In this case, the influence functional is expressed as
\begin{eqnarray}
 F[ \eta, \xi ] &=&
  {\rm exp} [ - i \frac{( \gamma_N \hbar A_{\perp})^2}{4}
 \int_{0}^{t} \int_{0}^{t_1} dt_1 dt_2  \nonumber \\  
 && \{ \xi (t_1) (  2 G^A (t_1,i,t_2,i) )
  \eta (t_2) \nonumber \\
 &+& \eta (t_1) ( 
 2 G^R (t_1,i,t_2,i) ) \xi (t_2) \nonumber \\
 &-& \eta (t_1) ( 
 2 G^R (t_1,i,t_2,i) ) \xi (t_2) \}
 ],
\end{eqnarray}
where $G^R (t_1,i,t_2,j)$, $G^A (t_1,i,t_2,j)$ 
 and $G^K (t_1,i,t_2,j)$ are
 retarded Green's function, advanced Green's function 
 and Keldysh Green's function.
For above integral equation, we slice the time and take a difference, we get the differential equation for density matrix is given by,
\begin{eqnarray}
&& \frac{ d \rho_{\rm b} (t)}{d t} = 
- \frac{i}{\hbar} [H_{qb}, \rho_{\rm b} (t)]
 -\frac{i}{4} (\gamma_N \hbar A_{\perp})^2 
\int_0^t dt_1\nonumber \\
&& \left(
 \begin{array}{cc}
  0 & 2 G^R (i, t,i, t_1)  \\
  2 G^A (i, t,i, t_1), 
  & - 2 G_K (i, t, i, t_1) \\ 
  \end{array}
  \right)  \rho_{\rm b} (t_1), 
\end{eqnarray}
where $\rho_{\rm b} (t)$ is
\begin{eqnarray}
\rho_{\rm b} (t) = \left( \begin{array}{cc}
  \eta_z (t) =1,  & \eta_z (t)= -1  \\
  \xi_z (t) =1, & \xi_z (t) = -1 \\ 
  \end{array} \right).
\end{eqnarray}  
Next we choose the representation of density matrix for the spin diagonal case,
\begin{eqnarray}
\label{eqn:rho}
&& \frac{ d \rho_{\bf s} (t)}{d t}
= - \frac{i}{\hbar} [ H_{qb}, \rho_{\rm s}(t)] 
- \frac{i}{4} ( \gamma_N \hbar A_{\perp})^2 \int_0^t dt_1
\nonumber \\
&& \left(
  \begin{array}{c} 
- G^K (i,t,i,t_1) \\
  G^R (i,t,i,t_1) + G^A (i,t,i,t_1) \\
 - i  ( G^A (i,t,i,t_1) - G^R (i,t,i,t_1)) \\
 G^K (i,t,i,t_1) \\
 \end{array}
 \right)^{t}  \rho_{\bf s} (t_1), \nonumber \\
\end{eqnarray}
where the density matrix, $\rho_{\rm s} (t)$ is represented as
\begin{eqnarray}
\rho_{\rm s} (t) &=& \left( \begin{array}{c} 
  |\uparrow (t) >< \uparrow (t)| + |\downarrow(t)><\downarrow(t)|,\\
  |\uparrow (t)><\downarrow (t)|,  \\
  |\downarrow (t) >< \uparrow (t) |, \\ 
  |\uparrow (t) >< \uparrow (t)|-|\downarrow(t)><\downarrow(t)|  \\ 
  \end{array} \right). \nonumber \\
\end{eqnarray}
The trace of density matrix decreases with time.
This event is decoherence.
For spin flip process, another diagonal element increases with time, this represents self-excitation.
One of off-diagonal element shows decoherence.
Another off-diagonal element shows oscillation that modulates the signal.

\section{one-dimensional qubit systems}

Next, we examine one-dimensional qubit systems by using influence functional.
\cite{Itakura}
In this case, the effect of indirect interaction appears.
The interaction Hamiltonian is as below,
\begin{equation}
  H_{int} = \gamma_N \hbar^2 \sum_i [  
  A_{\perp} ( I_{i}^+ c_{i} + I_{i}^- c_{i}^{\dagger})].
\end{equation}
We integrate about spin-less bath system and 
the qubit systems except the i-th site qubit.
This lead density matrix of qubit system,
\begin{eqnarray}
 \rho ( I_{zi+}^f, I_{zi-}^f )
   &=& \Pi_i \int^{I_{zi+}(t) =I_{zi+}^f, I_{zi-} (t) =
    I_{zi-}^f}_{ I_{zi+}(0)=I_{zi+}^f, I_{zi+} (0) = I_{zi+}^f} 
  [ d I_{zi+} ] [ d I_{zi-} ] \nonumber \\ 
 && {\exp} ( \frac{i}{\hbar} (I_{qb} [ I_{zi-} ] - I_{qb} [ I_{zi+} ] ))
  F[I_{zi+},I_{zi-}] \nonumber \\
\end{eqnarray}
the influence functional is
\begin{eqnarray}
 F[ I_{zi+}, I_{zj-} ] &=& \int \Pi_i 
 [ d c_{+i}^{\dagger}  ] [ d  c_{-i}^{\dagger} ]
 [ d c_{+i}  ] [ d  c_{-i} ] \nonumber \\
 && \delta (  c_{+i} (t) - c_{-i}(t) ) 
  \rho ( c_i  (0),  c_j (0) ) \nonumber \\
 && {\rm exp} \{ 
 \frac{i}{\hbar} (  I[c_+ ]- I [c_- ] ) 
\}. \nonumber \\
\end{eqnarray}
where, $I_{qb} [I] = \int_0^t \hbar \Delta I_{iz}$.
For following discussion, we choose the action of system,
$ I[ c ] = I_0 [ c ] + I_{int} [ I_{zi}, c ]$, thus the unperturbed is given by,
\begin{eqnarray}
   I_0 [ c ] &=& \frac{ (A_{\perp} \gamma_N \hbar)^2}{4} \nonumber \\
   && ( c_i^{\dagger} (t_1)
    {\bf \Delta}_{00p}^{+--1} ( t_1 , i , t_2 , j) 
    c_j ( t_2 ) \nonumber \\
    &+&  c_i (t_1)
    {\bf \Delta}_{00p}^{-+-1} ( t_1 , i , t_2 , j) 
    c_j^{\dagger} ( t_2 )), \nonumber \\ 
\end{eqnarray}
Introducing the incoming interaction picture for the environment system, 
the equation turns out,  
\begin{eqnarray}
  F[ I_{zi+}, I_{zi-}] &=& 
   {\rm exp} [ - i \frac{(\gamma_N \hbar A_{\perp})^2}{4} 
    \int_{0}^{t} \int_{0}^{t_1} dt_1 dt_2 \nonumber \\  
   &&  
   ( I_{i++} (t_1) \Delta_{00}^{++} (i,t_1,j,t_2) I_{j++} (t_2)  \nonumber \\
    &+& I_{i+-} (t_1) \Delta_{00}^{--} (i,t_1,j,t_2) I_{j+-} (t_2) \nonumber \\
    &-& I_{i++} (t_1) \Delta_{00}^{+-} (i,t_1,j,t_2) I_{j+-} (t_2) \nonumber \\
   &-&
    I_{i+-} (t_1) \Delta_{00}^{-+} (i,t_1,j,t_2) I_{j++} (t_2) ) \nonumber \\
  && 
   ( I_{i-+} (t_1) \Delta_{00}^{++} (i,t_1,j,t_2) I_{j-+} (t_2) \nonumber \\
    &+& I_{i--} (t_1) \Delta_{00}^{--} (i,t_1,j,t_2) I_{j--} (t_2) \nonumber \\
   &-&  I_{i-+} (t_1) \Delta_{00}^{+-} (i,t_1,j,t_2) I_{j--} (t_2) \nonumber \\
   &-& I_{i--} (t_1) \Delta_{00}^{-+} (i,t_1,j,t_2) I_{j-+} (t_2)) ].
    \nonumber \\   
\end{eqnarray}
For the convenience we change of the coordinate,
\begin{equation}
  \eta_{zi} = (I_{zi+} + I_{zi-})/2, 
  \xi_{zi} = (I_{zi+} - I_{zi-})/2.
\end{equation}
The $\eta$ and $\xi$ are called sojourn and blip.
In terms of this variable, density matrix is described by as follows
\begin{eqnarray}
 \rho (\eta_{zi} =1) &=& |\uparrow_i><\uparrow_i|,
 \rho ( \eta_{zi} =-1) = |\downarrow_i><\downarrow_i|, \nonumber \\
 \rho (\xi_{zi}  = 1 ) &=&  |\uparrow_i><\downarrow_i|,
 \rho ( \xi_{zi} = - 1)  = |\downarrow_i><\uparrow_i |. \nonumber 
\end{eqnarray}
Other variables are defined by
\begin{equation}
  \eta_{i+} = (I_{i++} + I_{i+-})/2, 
  \xi_{i+} = (I_{i++} - I_{i+-})/2,
\end{equation}
\begin{equation}
  \eta_{i-} = (I_{i-+} + I_{i--})/2, 
  \xi_{i-} = (I_{i-+} - I_{i--})/2.
\end{equation}
In terms of above variables the elements of density matrix are expressed as
\begin{eqnarray}
 \rho (\eta_{i+} =1) &=& |\uparrow_i><\downarrow_i|,
 \rho ( \eta_{i+} =-1) = |\downarrow_i><\uparrow_i|, \nonumber \\
 \rho (\xi_{i+}  = 1 ) &=&  |\uparrow_i><\uparrow_i|,
 \rho ( \xi_{i+} = - 1)  = |\downarrow_i><\downarrow_i |. \nonumber \\
 \rho (\eta_{i-} =1 ) &=& |\downarrow_i><\uparrow_i|,
 \rho ( \eta_{i-} =-1) = |\uparrow_i><\downarrow_i|, \nonumber \\
 \rho (\xi_{i-}  = 1 ) &=&  |\downarrow_i><\downarrow_i|,
 \rho ( \xi_{i-} = - 1)  = |\uparrow_i><\uparrow_i |. \nonumber 
\end{eqnarray}
Therefore, the below equations are hold for these new variables,
\begin{equation}
\eta_{zi} = \xi_{i+} = -\xi_{i-} , \xi_{zi} = \eta_{i+} =-\eta_{i-}.
\end{equation}

In this case, the influence function is expressed as
\begin{eqnarray}
 && F[ \eta_i, \xi_i ] =
  {\rm exp} [ - i \frac{( \gamma_N \hbar A_{\perp} )^2}{4}
 \int_{0}^{t} \int_{0}^{t_1} dt_1 dt_2 \nonumber \\  
 && \{ \xi_{zi} (t_1) ( 2 G^A (t_1,i,t_2,j) )
  \eta_{zj} (t_2) \nonumber \\
 &+& \eta_{zi} (t_1) (2 G^R (t_1,i,t_2,j) ) \xi_{zj} (t_2) \nonumber \\
 &-& \xi_{zi} (t_1) ( 2 G^R (t_1,i,t_2,j) ) \xi_{zj} (t_2) \} ],
  \nonumber \\
\end{eqnarray}
where $G^R (t_1,i,t_2,j)$, $G^A (t_1,i,t_2,j)$ 
 and $G^K (t_1,i,t_2,j)$ are
 retarded Green's function, advanced Green's function 
 and Keldysh Green's function.
For above integral equation, we slice the time and take difference, we get the differential equation for density matrix is given by,
\begin{eqnarray}
&&\frac{ d \rho_{\rm b}  (i,j,t)}{d t} = 
- \frac{i}{\hbar} [H_{qb}, \rho_{\rm b}(t)] 
-\frac{i}{4} (\gamma_N \hbar A_{\perp})^2 
\int_0^t dt_1 \sum_k  \nonumber \\
&& \left(
 \begin{array}{cc}
  0 & 2 G^R (i, t,k, t_1)  \\
  2 G^A (i, t,k, t_1) 
  & - 2 G_K (i, t, k, t_1) \\ 
  \end{array}
  \right) \rho_{\rm b} (t_1,k,j). \nonumber \\
\end{eqnarray}
Next we choose the representation of density matrix for the spin diagonal case,
\begin{eqnarray}
&& \frac{ d \rho_{\rm s} (i,j,t)}{d t} = 
 - \frac{i}{\hbar}
  [H_{qb}, \rho_{\rm s} (t)] - \frac{i}{4} ( \gamma_N \hbar A_{\perp} )^2 \int_0^t dt_1 \sum_k
 \nonumber \\
&& \left(
  \begin{array}{c} 
- G^K (i,t,k,t_1) \\
  G^R (i,t,k,t_1) + G^A (i,t,k,t_1) \\
 - i   
  ( G^A (i,t,k,t_1) - G^R (i,t,k,t_1)) \\
 G^K (i,t,k,t_1) \\
 \end{array}
 \right) \rho_{\rm s} (k,j,t_1). \nonumber \\
\end{eqnarray}
The Green's functions of free bath system are given by
\cite{Zagoskin}
\begin{eqnarray}
G^R (k, w) &=& \frac{1}{w-(\omega (k) - \mu) + i 0}, \nonumber \\
G^A (k, w) &=& \frac{1}{w-(\omega (k) - \mu) - i 0}, \nonumber \\
G^k (k, w) &=& - 2 \pi i (1 \mp 2 n_k) \delta ( w - (\omega (k) - \mu)),
 \nonumber \\
n_k &=& \frac{1}{e^{\frac{\hbar \omega (k)}{k_B T}} \pm 1},
\end{eqnarray}
where up sign is fermion and down sign is boson,
$\mu$ is chemical potential.
The trace of density matrix decreases with time.
For qubit flip process, another diagonal element increases with time, this represents self-excitation. 
One of off-diagonal element shows decoherence.
Another off-diagonal element shows oscillation that modulates the signal.
  
\section{multi-band one-dimensional system}

Next, we examine multi-band qubit systems by using influence functional.
\cite{Itakura}
The interaction Hamiltonian is as below,
\begin{equation}
  H_{int} = \gamma_N \hbar^2 \sum_{i,\sigma,m}^{\uparrow \downarrow 2} [  
  A_{\perp} ( I_{i}^+ c^m_{i,\sigma} + I_{i}^- c_{i,\sigma}^{\dagger,m} ) ].
\end{equation}
We integrate about the qubit system except the i-th site qubit.
m is multi-band index.
This lead density matrix of qubit system,
\begin{eqnarray}
 \rho ( I_{zi+}^f, I_{zi-}^f )
   &=&  \int^{I_{zi+}(t) =I_{zi+}^f, I_{zi-} (t) =
    I_{zi-}^f}_{ I_{zi+}(0)=I_{zi+}^f, I_{zi+} (0) = I_{zi+}^f} 
 \Pi_i [ d I_{zi+} ] [ d I_{zi-} ] \nonumber \\ 
 && {\exp} ( \frac{i}{\hbar} (I_{qb} [ I_{zi-} ] - I_{qb} [ I_{zi+} ] ))
  F[I_{zi+},I_{zi-}].
\end{eqnarray}
The influence functional is
\begin{eqnarray}
 F[ I_{zi+}, I_{zj-} ] &=& \int \Pi_{i,\sigma,m}
 [ d c_{+i,\sigma}^{\dagger.m}  ]
  [ d  c_{-i,\sigma}^{\dagger,m} ]
 [ d c_{+i,\sigma}^m  ] [ d  c_{-i,\sigma}^m ] \nonumber \\
 && \delta (  c_{+i,\sigma}^m (t) - c_{-i,\sigma}^{m}(t) ) 
  \rho ( c_{i,\sigma}^m  (0),  c_{j,\sigma}^m (0) ) \nonumber \\
 && {\rm exp} \{ 
 \frac{i}{\hbar} (  I[c_{+,\sigma}^m ]- I [c_{-,\sigma}^m ] ) 
\}. \nonumber \\
\end{eqnarray}
where, $I_{qb} [I] = \int_0^t \hbar \Delta I_{iz}$.
For following discussion, we choose the action of system,
$ I[ c_{\alpha} ] = I_0 [ c_{\alpha} ] + I_{int} [ I_{zi}, c_{\alpha} ]$, 
thus the unperturbed is given by,
\begin{eqnarray}
   I_0 [ c_{\sigma}^m ] &=& 
   \frac{ (A_{\perp} \gamma_N \hbar)^2}{4} \nonumber \\
   && ( c_{i \sigma}^{\dagger,m} (t_1)
    {\bf \Delta}_{00p,\sigma m \sigma' m'}^{+--1} ( t_1 , i , t_2 , j) 
    c_{j \sigma'}^{m'} ( t_2 ) \nonumber \\
    &+&  c_{i \sigma}^m (t_1)
    {\bf \Delta}_{00p,\sigma m \sigma' m'}^{-+-1} ( t_1 , i , t_2 , j) 
    c_{j \sigma'}^{\dagger m'} ( t_2 )), \nonumber \\ 
\end{eqnarray}
Introducing the incoming interaction picture for the environment system,
the influence function is expressed as
\begin{eqnarray}
 && F[ \eta_i, \xi_i ] =
  {\rm exp} [ - i \frac{( \gamma_N \hbar A_{\perp} )^2}{4}
 \int_{0}^{t} \int_{0}^{t_1} dt_1 dt_2 \sum_{\sigma,m} \nonumber \\
 && \{ \eta_i (t_1)  G^R (t_1,i,\sigma,m,t_2,j,\sigma,m)
  \xi_j (t_2) \nonumber \\ 
 &+& \xi_i (t_1) G^A (t_1,i,\sigma,m,t_2,j,\sigma,m) \eta_j (t_2) \nonumber \\
                     &-& \eta_i (t_1) G^K (t_1,i,\sigma,m,t_2,j,\sigma,m) 
                      \eta_j (t_2) \} \nonumber \\  
 &+& \{ \eta_i (t_1)  G^R (t_1,i,\sigma,m,t_2,j,\sigma,m) 
 \xi_j (t_2) \nonumber \\
 &+& \xi_i (t_1) G^A (t_1,i,\sigma,m,t_2,j,\sigma,m)
  \eta_j (t_2) \nonumber \\
                      &-& \eta_i (t_1) G^K (t_1,i,\sigma,m,t_2,j,\sigma,m) 
                      \eta_j (t_2) \}]   \nonumber \\
&=&
  {\rm exp} [ - i \frac{( \gamma_N \hbar A_{\perp} )^2}{4}
 \int_{0}^{t} \int_{0}^{t_1} dt_1 dt_2 \sum_{\sigma m}\nonumber \\  
 && \{ \xi_i (t_1) ( 2 G^A (t_1,i,\sigma,m,t_2,j,\sigma,m) )
  \eta_j (t_2) \nonumber \\
 &+& \eta_i (t_1) (2 G^R (t_1,i,\sigma,m,t_2,j,\sigma,m) ) \xi_j (t_2) \nonumber \\
 &-& \xi_i (t_1) ( 2 G^R (t_1,i,\sigma,m,t_2,j,\sigma,m) ) \xi_j (t_2) \} ],
  \nonumber \\
\end{eqnarray}
where $G^R (t_1,i,\sigma,m,t_2,j,\sigma,m)$,
 $G^A (t_1,i,\sigma,m,t_2,j,\sigma,m)$ 
 and $G^K (t_1,i,\sigma,m,t_2,j,\sigma,m)$ are
 retarded Green's function, advanced Green's function 
 and Keldysh Green's function.
For above integral equation, we slice the time and take difference, we get the differential equation for density matrix for the spin diagonal case,
\begin{eqnarray}
&& \frac{ d \rho_{\rm s} (i,j,t)}{d t} = 
 - \frac{i}{\hbar}
  [H_{qb}, \rho_{\rm s} (t)] - \frac{i}{4} ( \gamma_N \hbar A_{\perp} )^2 \int_0^t dt_1 \sum_{k,\sigma,m}
 \nonumber \\
&& \left(
  \begin{array}{c} 
- G^K (i,t,\sigma,m,k,t_1,\sigma,m) \\
  G^R (i,t,\sigma,m,k,t_1,\sigma,m) + G^A (i,t,\sigma,m,k,t_1,\sigma,m) \\
 - i   
  ( G^A (i,t,\sigma,m,k,t_1,\sigma,m) - G^R (i,t,\sigma,m,k,t_1,\sigma,m)) \\
 G^K (i,t,\sigma,m,k,t_1,\sigma,m) \\
 \end{array}
 \right) \nonumber \\
 &&\rho_{\rm s} (k,j,t_1). \nonumber \\
\end{eqnarray}
Even if multi-band system, we only need diagonal band Green's functions.
The Green's functions are obtained.
\cite{Itakura_Kawakami}
The trace of density matrix decreases with time.
For qubit flip process, another diagonal element increases with time, this represents self-excitation. 
One of off-diagonal element shows decoherence.
Another off-diagonal element shows oscillation that modulates the signal.

The von Neumann entropy is given by following equation.
\begin{equation}
S = - {\rm Tr} \rho  \log \rho  = - \lim_{n \rightarrow 1} 
\frac{\partial}{\partial n} \rho ^n.
\end{equation}
The time evolution of the entropy is obtained as below,
\begin{eqnarray}
\frac{d S}{d t} &=&
   \frac{i}{4} ( \gamma_N \hbar A_{\perp} )^2 \int_0^t dt_1 \sum_{i,j,\sigma,m}
 \nonumber \\
&&  
{\rm Tr} [ \rho_{\rm s} (j,i,t_1)
\nonumber \\
&& (1 + \log ( \rho_{\rm s} (j,i,t_1)) ] .
\end{eqnarray}

\section{numerical results} 

\begin{figure}[tb]
\begin{center}
\unitlength 1mm
\begin{picture}(90,90)(0,5)
\put(0,0){\resizebox{60mm}{!}{\includegraphics{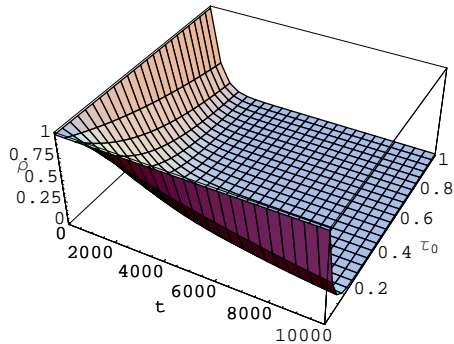}}}
\end{picture}
\end{center}
\caption{\label{fig:exact} Time dependence of trace for 
$(\gamma_N \hbar A_{\perp} J_C)^2$=0.01, 
$\rho^{|\uparrow><\uparrow|+|\downarrow><\downarrow|}(0) = 1$
One axis is time and another axis is time constant.}
\end{figure}
\begin{figure}[tb]
\begin{center}
\unitlength 1mm
\begin{picture}(90,90)(0,5)
\put(0,0){\resizebox{60mm}{!}{\includegraphics{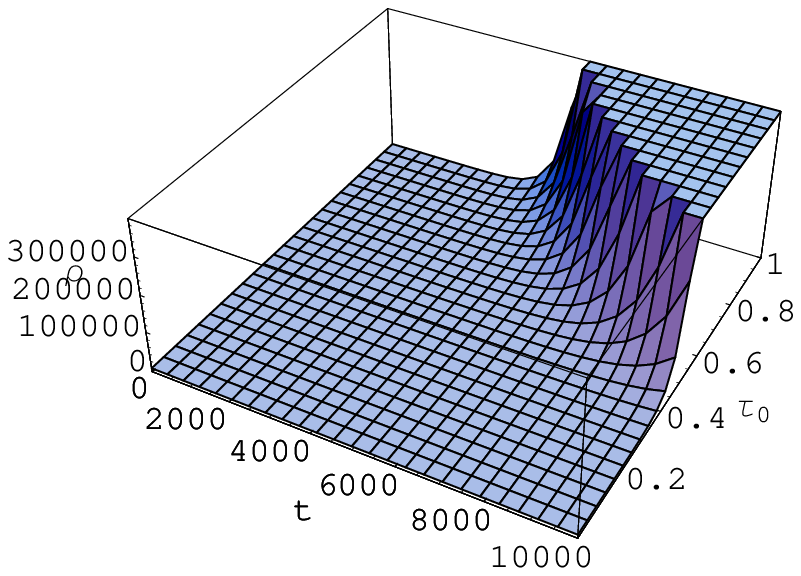}}}
\end{picture}
\end{center}
\caption{\label{fig:exact} Time dependence of another diagonal element for 
$(\gamma_N \hbar A_{\perp} J_C)^2$=0.01, 
$\rho^{|\uparrow><\uparrow|-|\downarrow><\downarrow|}(0) = 1$.}
\end{figure}
\begin{figure}[tb]
\begin{center}
\unitlength 1mm
\begin{picture}(90,90)(0,5)
\put(0,0){\resizebox{60mm}{!}{\includegraphics{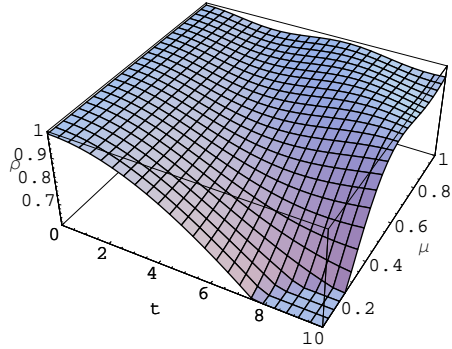}}}
\end{picture}
\end{center}
\caption{\label{fig:exact} Time dependence of trace for 
$(\gamma_N \hbar A_{\perp} )^2$=0.01, 
$\rho^{|\uparrow><\uparrow|+|\downarrow><\downarrow|}(0) = 1$
One axis is time and another axis is chemical potential.}
\end{figure}
\begin{figure}[tb]
\begin{center}
\unitlength 1mm
\begin{picture}(90,90)(0,5)
\put(0,0){\resizebox{60mm}{!}{\includegraphics{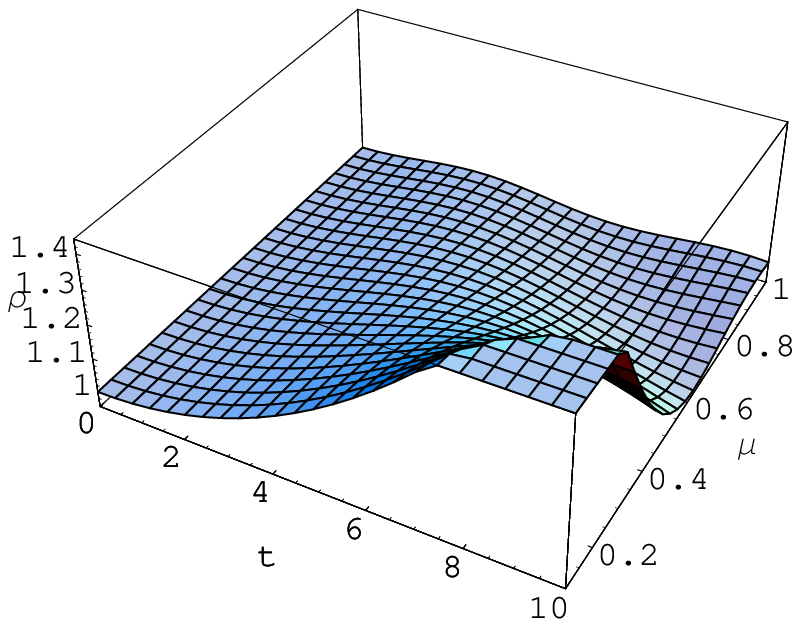}}}
\end{picture}
\end{center}
\caption{\label{fig:exact} Time dependence of another diagonal element for 
$(\gamma_N \hbar A_{\perp})^2$=0.01, 
$\rho^{|\uparrow><\uparrow|-|\downarrow><\downarrow|}(0) = 1$.}
\end{figure}
\begin{figure}[tb]
\begin{center}
\unitlength 1mm
\begin{picture}(90,90)(0,5)
\put(0,0){\resizebox{60mm}{!}{\includegraphics{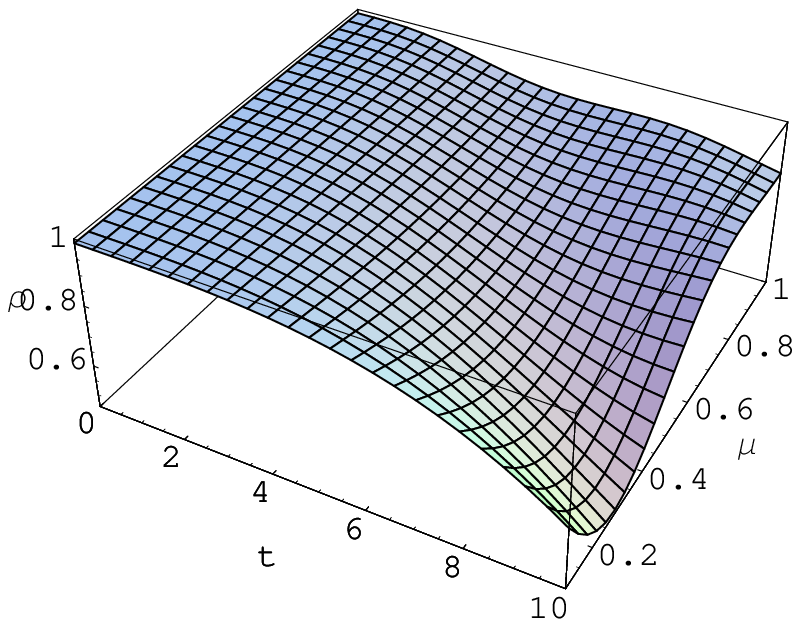}}}
\end{picture}
\end{center}
\caption{\label{fig:exact} Time dependence of off-diagonal element for 
$(\gamma_N \hbar A_{\perp})^2$=0.01, 
$\rho^{|\uparrow><\downarrow|.}(0) = 1$.}
\end{figure}
\begin{figure}[tb]
\begin{center}
\unitlength 1mm
\begin{picture}(90,90)(0,5)
\put(0,0){\resizebox{60mm}{!}{\includegraphics{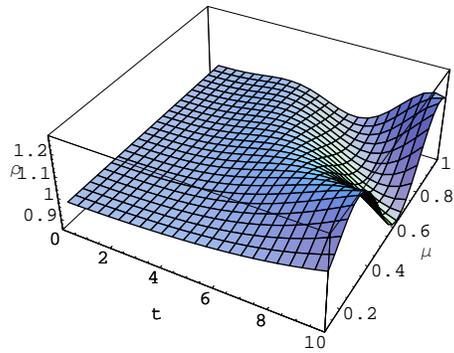}}}
\end{picture}
\end{center}
\caption{\label{fig:exact} Time dependence of another off-digonal element for 
$(\gamma_N \hbar A_{\perp} )^2$=0.01, 
$\rho^{|\downarrow><\uparrow|.}(0) = 1$.}
\end{figure}
\begin{figure}[tb]
\begin{center}
\unitlength 1mm
\begin{picture}(90,90)(0,5)
\put(0,0){\resizebox{60mm}{!}{\includegraphics{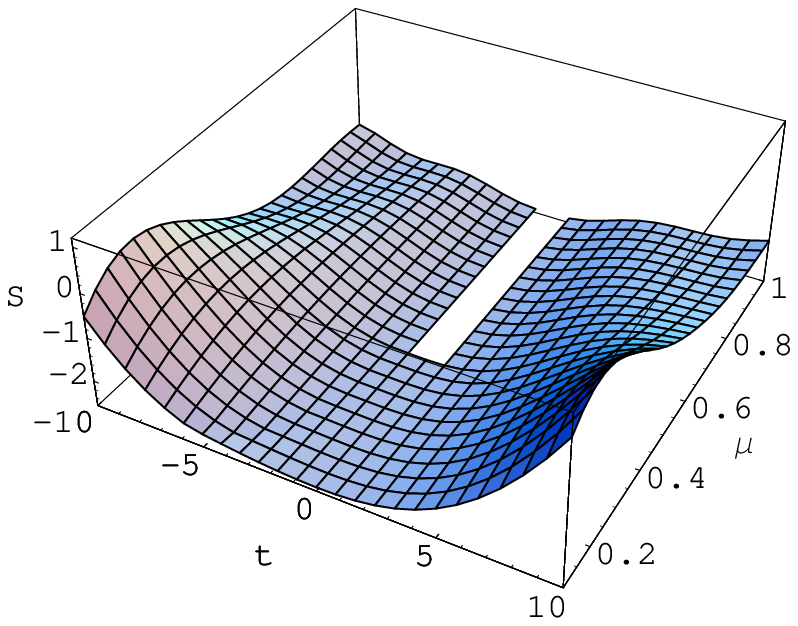}}}
\end{picture}
\end{center}
\caption{\label{fig:exact} Time dependence of von Neumann entropy for 
$(\gamma_N \hbar A_{\perp} )^2$=0.01, 
$\rho_s (0) = (1, 1;1, 1)$.}
\end{figure}

Here we present the numerical calculation of the density matrix and entropy.

When environment is random telegraph noise system,
 the analytical expression is 
$G^K (t) = i J_C e^{-\frac{|t|}{\tau_0}}$.
The results show with increasing time constant the trace decreases
and another diagonal element increases. 
This indicates that the slower random telegraph noise leads to the more mixed state.
Thus the background charge fluctuation is important for charge qubit.
\cite{Itakura_Tokura}

Next, we examine the 0-dimensional free fermion bath.
The density matrix under rotating wave approximation, four qubit density matrix are given by Fig. 3, 4, 5 and 6.
These results show analytical described behavior.
The von Neumann entropy also calculated as Fig 7.
The entropy increases with time increases.
This quantity is symmetric about origin of time 
and oscillates with changing the chemical potential, because
 the at t=0 time qubit and environment does not entangle.
It should be noted that present numerical estimations are rotating wave approximation.
Thus this numerical study is correct only short time regime.

\section{Conclusion}

In summary, we had examined decoherence of qubit systems coupled with one-dimensional bath.
The examined decoherence is the case of single qubit system and one-dimensional qubit systems.
We obtained the differential-integral equation for general initial condition.
The trace of density matrix decreases with time.
Another diagonal element increases with time, this represents self-excitation. 
One of off-diagonal element shows decoherence.
Another off-diagonal element shows oscillation that modulates the signal.
Decoherence without trace conservation occurs. 
We obtained the time evolution of von Neumann entropy.
We also present the numerical calculation of density matrix and the entropy.
For the random telegraph noise,
the results indicate that the slower random telegraph noise leads to the more mixed states.
Thus the background charge fluctuation is important for charge qubit.


\end{document}